Перспективные люминофоры для белых светодиодов на основе двойных молибдатов состава $Ln_2Zr_3(MoO_4)_9$ (Ln: Eu, Tb).

Д. Софич[1], С. Доржиева[2], О. Чимитова[2], Б.Г. Базаров[2], Ю.Л. Тушинова[2], Ж.Г. Базарова[2], Р. Шендрик[1]

1) Институт геохимии имени А.П. Виноградова Сибирского отделения Российской академии наук, 664033, г. Иркутск, Россия.

2) Байкальский институт природопользования Сибирского отделения Российской академии наук.

E-mail: *sofich-dmitriy@live.com*

*В статье приводятся результаты исследования спектральных характеристик новых люминофоров на основе редкоземельных ионов $Eu^{3+}$ и $Tb^{3+}$ в матрицах двойных молибдатов. Рассмотрены спектры люминесценции и возбуждения, определены времена затухания основных переходов, определены цветовые координаты. Результаты показывают возможную пригодность использования данных соединений в качестве компонентов люминофоров белых светодиодов.*

В настоящее время осветительные приборы на основе белых светодиодов практически вытеснили с рынка бытового освещения газоразрядные лампы, а также, лампы накаливания. Лампы на основе белых светодиодов намного экологичнее, надежнее и экономичнее их предшественников. На текущем этапе развития светодиодных технологий наиболее актуальными задачами являются улучшение вывода света, увеличение коэффициента цветопередачи и величины квантового выхода светодиодов[1]. Для увеличения коэффициента цветопередачи в люминофор часто вводятся примеси, призванные дополнить спектр свечения в нужной области. Для таких примесей предъявляются повышенные требования к длине волны возбуждения, цветовым координатам их свечения, оптическому поглощению, химической и термической стабильности. Для уменьшения цветовой температуры светодиодов часто применяются красные люминофоры с трехвалентным европием[2].

В данной работе изучаются синтезированные мелкозернистые порошки двойных молибдатов следующих составов: $Tb_2Zr_3(MoO_4)_9$ и $Eu_2Zr_3(MoO_4)_9$. Порошки синтезировались по керамической технологии путем ступенчатого отжига смеси стехиометрических количеств $Eu_2O_3(Tb_2O_3)$, $MoO_3$ и $ZrO_2$ в течение 150 часов до максимальной температуры в 700ºС[3]. Спектры люминесценции были зарегистрированы при помощи двойного монохроматора СДЛ-1 с решетками 600 lines/mm и фотоэлектронного умножителя ФЭУ-106, возбуждение производилось при помощи ксеноновой лампы высокого давления ДКШ-150 через монохроматор МДР-2 с дифракционной решеткой 1200 lines/mm. В качестве подложки для нанесения образца выступал полированный беспримесный кристалл LiF, который закреплялся в держателе. После измерений спектры возбуждения корректировались на спектр возбуждения родамина-6ж. Измерения времен затухания люминесценции проводились при возбуждении импульсами аргоновой лампы, совмещенной с монохроматором МДР-2. К приемнику подключался двухканальный цифровой осциллограф Rigol DS1102E. Кривые затухания регистрировались в режиме накопления.

На рисунке 1 показаны спектры возбуждения и люминесценции $Tb_2Zr_3(MoO_4)_9$, наблюдаются интенсивные полосы f-f свечения в области 480-680 nm, являющиеся переходами с уровня $^5D_4$ на нижележащие уровни $^7F_{0,1,2,3,4,5,6}$. Самой интенсивной является полоса $^5D_4$-$^7F_5$ (543 nm). f-f полосы очень узкие, с полушириной 3-5 nm, это связано с тем, что 4f- оболочка экранирована вышележащими 5s и 5p электронными оболочками $Tb^{3+}$. В спектре возбуждения присутствуют полосы, связанные с f-f переходами с основного состояния $^7F_6$, при возбуждении в f- полосы наибольшая интенсивность свечения достигается при длине волны возбуждения 380 nm (переход $^7F_6$-$^5D_3$). Также, в области 300 nm присутствует широкая интенсивная полоса возбуждения. Измерения времен затухания основных переходов показали, что характерное время затухания люминесценции составляет 420 μs, при этом, формы кривых соответствуют моноэкспоненциальному закону. Такие времена характерны для излучательных f-f переходов $Tb^{3+}$ в шеелитоподобных матрицах[4].

По спектру люминесценции были определены цветовые координаты в пространстве CIE 1931: x=0,36 y=0,51 (Рис. 3). Цвет люминесценции — зеленый, т.к. основной излучающий переход — $^5D_4$-$^7F_5$ (543 nm), но, имеется уклон в белый цвет из-за наличия переходов $^5D_4$-$^7F_6$ (490 nm), $^5D_4$-$^7F_4$ (582 nm) и $^5D_4$-$^7F_3$.(618 nm).

Таким образом, $Tb_2Zr_3(MoO_4)_9$ - достаточно интересный зеленый люминофор с интенсивным возбуждением в области ближнего и среднего ультрафиолета, обладает интенсивной зеленой люминесценцией, обусловленной f-f переходами иона $Tb^{3+}$. По кривым затухания можно судить о том, что, несмотря на высокую концентрацию, ионы тербия слабо взаимодействуют друг с другом, не производя концентрационного тушения.

Спектр люминесценции $Eu_2Zr_3(MoO_4)_9$ (рис.2) содержит набор полос в красной области спектра (525-700 nm), главный переход - гиперчувствительный $^5D_0$-$^7F_2$ (615 nm). Присутствует переход $^5D_0$-$^7F_0$ (583 nm), наблюдающийся только при низкой симметрии окружения РЗИ[5]. По количеству пиков в каждом электрическом дипольном переходе группы $^7F_{0,1,2,3,4}$ была определена группа симметрии $C_4$ окружения РЗИ[6]. В спектре возбуждения имеются f-f переходы с основного f-уровня $^7F_0$, среди которых стоит выделить переходы $^7F_0$-$^5L_6$ (395 nm) и $^7F_0$-$^5D_2$ (464 nm), имеющие «удобные» длины волн для возбуждения существующими светодиодами, например синими или фиолетовыми чипами InGaN[7, 8]. Также, в области 300 nm имеется широкая полоса с эффективным возбуждением, совпадающая по форме и положению с полосой в $Tb_2Zr_3(MoO_4)_9$. Независимость положения и формы полосы от редкоземельного иона, входящего в состав соединения, говорит о том, что данная полоса скорее относится к переходу кислород — молибден, с последующим возбуждением f-f свечения редкоземельного иона. Вышесказанное свидетельствует о наличии в данных соединениях эффективного механизма передачи энергии от кристаллической матрицы к редкоземельному иону [9]. Затухание люминесценции основных переходов моноэкспоненциально, с характерным временем 400 µs. Определены цветовые координаты в пространстве CIE 1931: x=0.64 y=0.36. Цвет люминесценции — ярко-красный, с небольшим отклонением в оранжевый. Исходя из результатов исследований, авторы считают $Eu_2Zr_3(MoO_4)_9$ перспективным красным люминофором за счет интенсивной f-f люминесценции ионов $Eu^{3+}$. Этому сопутствует отсутствие концентрационного тушения при высокой концентрации ионов европия в кристаллической матрице, а положение полос в спектре возбуждения, как уже было сказано, достаточно удобно для возбуждения популярными синими и фиолетовыми InGaN светодиодами.

Также, возможно создание смеси из данных люминофоров, так, на рисунке 3 показаны цветовые координаты комбинированного спектра люминесценции $Eu_2Zr_3(MoO_4)_9$ и $Tb_2Zr_3(MoO_4)_9$, где максимальная интенсивность каждого спектра нормировалась на единицу. Кривая построена по промежуточным точкам, которые были вычислены путем изменения нормировочных множителей. Как видно из рисунка, смесь двух люминофоров дает желтый цвет, который можно сдвигать ближе к красному или зеленому, просто меняя соотношение между концентрациями люминофоров. Белый светодиод можно получить путем добавления синего люминофора при возбуждении фиолетовым светодиодом, т.к. при использовании синего светодиода не будет возбуждаться люминесценция в $Tb_2Zr_3(MoO_4)_9$.

Получены спектры люминесценции с набором узких пиков, относящихся к переходам внутри f-оболочки РЗИ, положение пиков было соотнесено с известными энергиями переходов и было установлено, что ионы Tb и Eu имеют валентность 3+. Исходя из данных исследований, можно сделать выводы о том, что в изученных соединениях редкоземельные ионы входят в кристаллическую матрицу равномерно и на большом расстоянии друг от друга, что позволяет избежать концентрационного тушения люминесценции, и, при этом, воздействие кристаллического окружения разрешает переходы внутри f-оболочки РЗИ. Наличие широкой полосы в спектрах возбуждения говорит о том, что в данных молибдатах имеется эффективный механизм передачи энергии с кристаллической матрицы на РЗИ вследствие переходов кислород-молибден.

Приведенные в работе результаты показывают возможность применения объектов исследования в качестве монохромных люминофоров с возбуждением от ультрафиолетовых, фиолетовых или синих светодиодов в случае $Eu_2Zr_3(MoO_4)_9$. Также, есть возможность добавить к смеси синий люминофор для получения белого светодиода с возможностью изменять его цвет на этапе приготовления люминофорного состава.

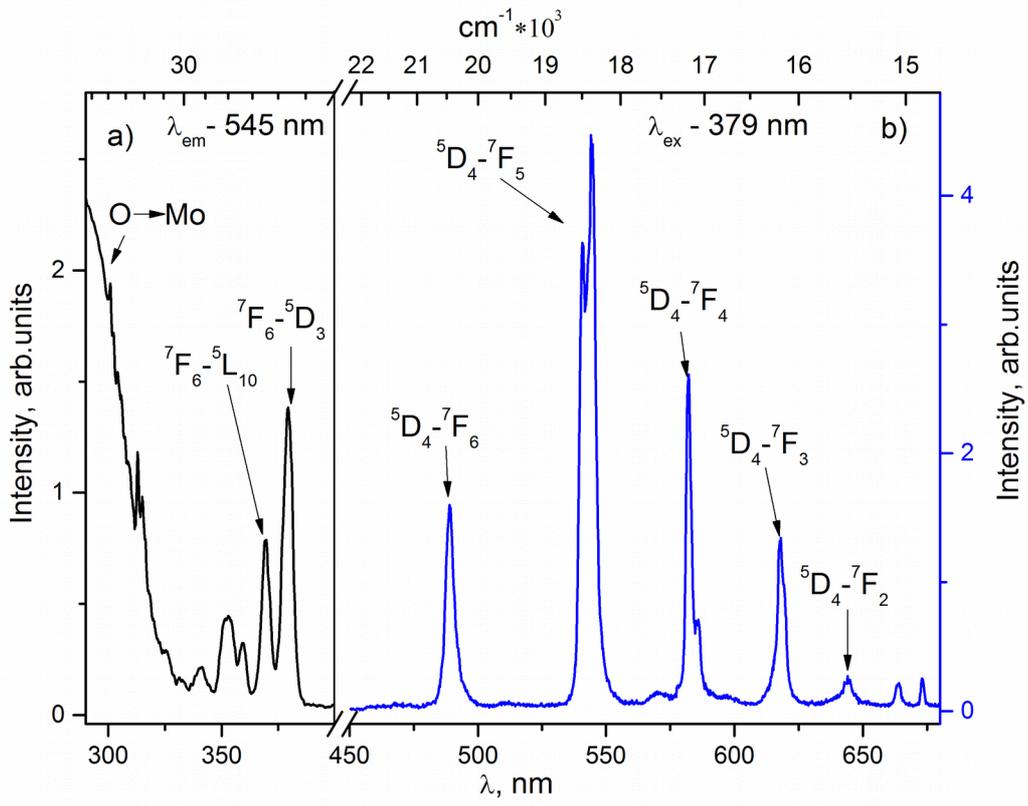

Рисунок 1.

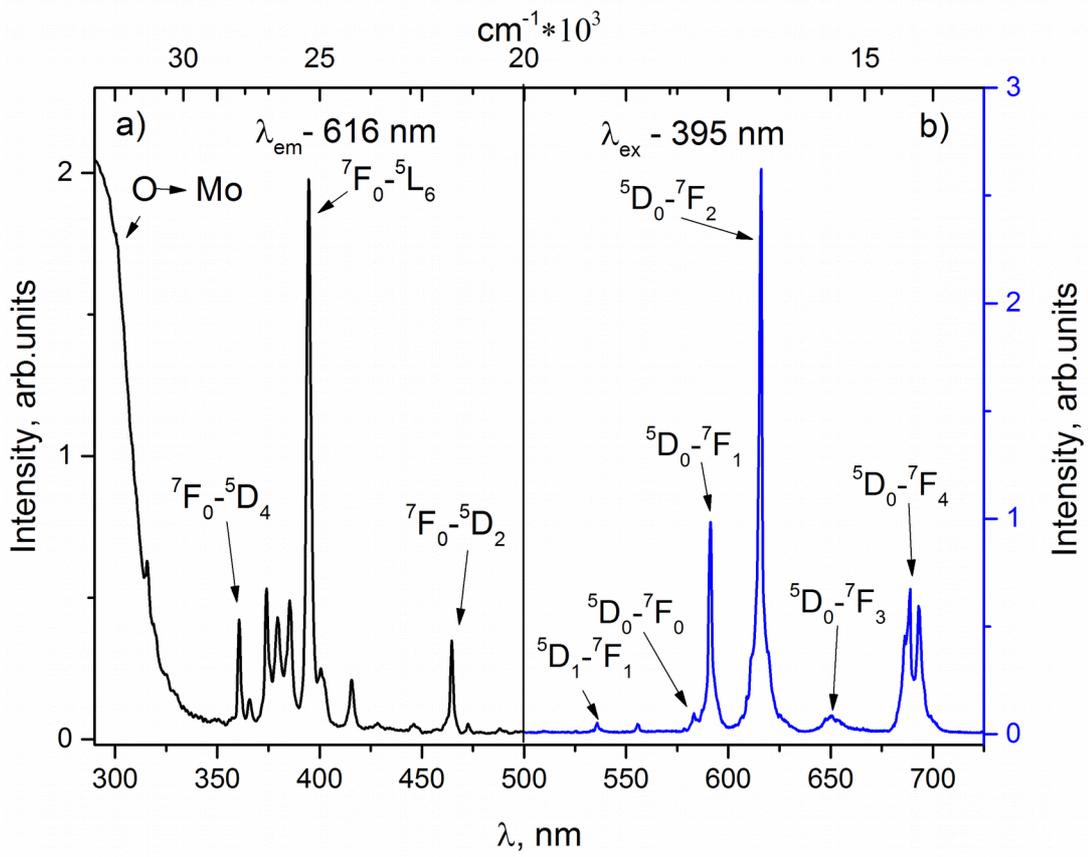

Рисунок 2.

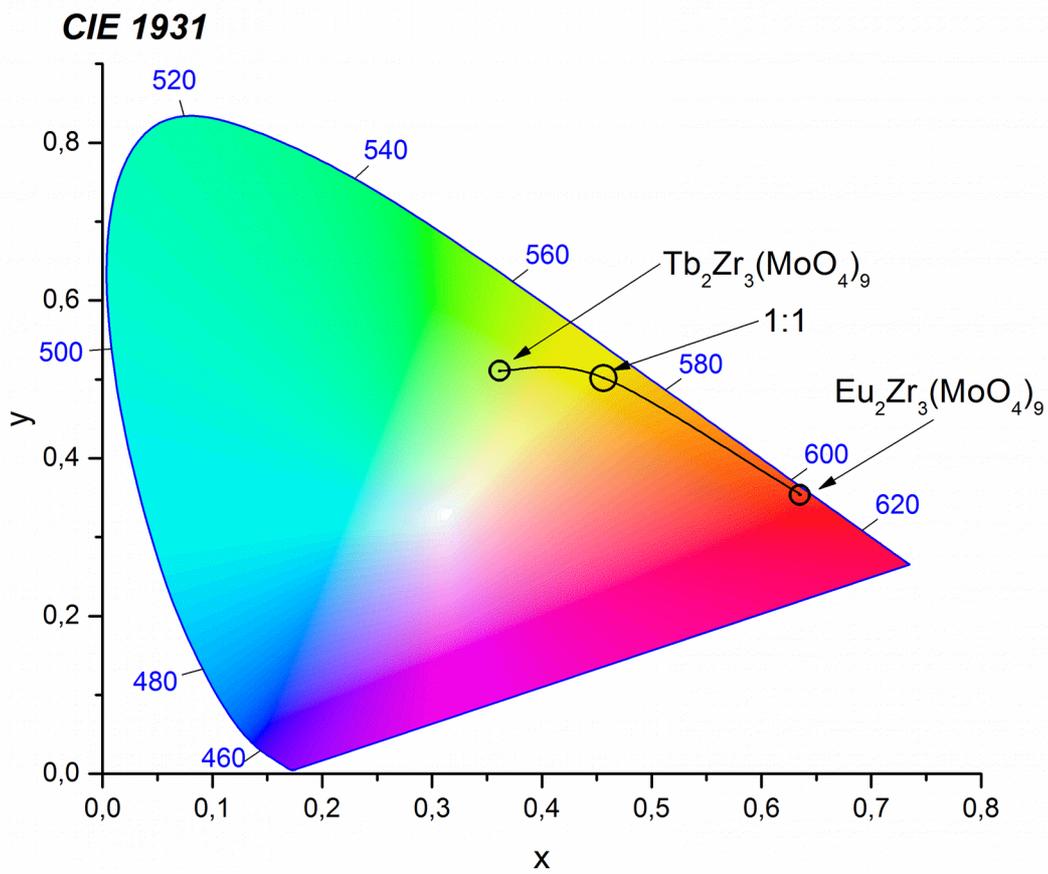

Рисунок 3.

Подписи к рисункам.

Рисунок 1. Спектры возбуждения (a) и свечения (b) $Tb_2Zr_3(MoO_4)_9$

Рисунок 2. Спектры возбуждения (a) и свечения (b) $Eu_2Zr_3(MoO_4)_9$

Рисунок 3. Цветовое пространство CIE 1931 с точками, соответствующими свечениям $Eu_2Zr_3(MoO_4)_9$ и $Tb_2Zr_3(MoO_4)_9$, а также, точкой «1:1» отмечен цвет смеси люминофоров при нормировании интенсивностей на 1.